# On the time to reach maximum for a variety of constrained Brownian motions

Satya N Majumdar and Julien Randon-Furling

*Laboratoire de Physique Théorique et Modèles Statistiques, Université Paris-Sud.*
*Bât. 100, 91405, Orsay Cedex, France*

Michael J Kearney

*Faculty of Engineering and Physical Sciences, University of Surrey,*
*Guildford, Surrey, GU2 7XH, United Kingdom*

Marc Yor

*Laboratoire de Probabilités et Modèles Aléatoires, Université Paris VI et VII,*
*4 Place Jussieu-Case 188, F-75252, Paris Cedex 05, France*

*Abstract*

We derive $P(M, t_m)$, the joint probability density of the maximum $M$ and the time $t_m$ at which this maximum is achieved, for a class of constrained Brownian motions. In particular, we provide explicit results for excursions, meanders and reflected bridges associated with Brownian motion. By subsequently integrating over $M$, the marginal density $P(t_m)$ is obtained in each case in the form of a doubly infinite series. For the excursion and meander, we analyse the moments and asymptotic limits of $P(t_m)$ in some detail and show that the theoretical results are in excellent accord with numerical simulations. Our primary method of derivation is based on a path integral technique; however, an alternative approach is also outlined which is founded on certain 'agreement formulae' that are encountered more generally in probabilistic studies of Brownian motion processes.

# 1. Introduction

Brownian motion (the Wiener process) is the most important and widely studied continuous-time stochastic process and, as such, has generated a huge literature. Despite this attention, however, it is still possible to identify problems relating to Brownian motion which are relatively easy to pose but not that well understood. Such problems are often directly linked to areas of application in the physical or social sciences, wherein their solution is of immediate relevance.

Within this overall context, there has been a recent renewal of interest in studying functionals of *constrained* Brownian motion. This has been driven by problems which arise quite naturally in, e.g., financial transactions [1], data storage in computer science [2], queueing dynamics [3], interface fluctuations [4] and extreme statistics in time series analysis [5]. For an overview, see [6–8]. The question we seek to answer in this paper is similarly motivated and finds its roots in the following classic problem. Given a Brownian motion, $x(\tau)$, on the interval $[0,t]$, subject to $x(0)=0$ but otherwise unconstrained, at what time, $t_m$, does $x(\tau)$ reach its maximum value, $M$? More precisely, what is the probability density, $P(t_m)$, associated with $t_m$? The answer is $P(t_m) = \frac{1}{\pi} t_m^{-1/2} (t-t_m)^{-1/2}$, or equivalently the cumulative distribution is given by $\Pr(t_m \leq x) = \frac{2}{\pi} \sin^{-1}[\sqrt{x/t}]$. This is Lévy's famous 'arcsine law' [9]. It is somewhat counterintuitive; the density has a minimum at the midpoint $t_m = t/2$ and peaks at the end points $t_m = 0$ and $t_m = t$, showing that Brownian motion is inherently 'stiff' [10,11]. The corresponding result for a Brownian bridge (which has the additional constraint that $x(t)=0$) is also known, namely $P(t_m) = 1/t$, which one may call the 'uniform law' [10]. Recently, an expression for $P(t_m)$ for a Brownian motion up to its first-passage time was also presented [12], adding to the results on first-passage Brownian functionals given in [13,14]. The main focus of the present paper is to derive $P(t_m)$ for three other cases: (i) a Brownian excursion, (ii) a Brownian meander, and (iii) a reflected Brownian bridge. We include the latter since it appears naturally in the context of establishing certain probability laws related to the excursion and the meander (see the discussion in section 4).

A Brownian excursion on the interval $[0, t]$ is defined as a Brownian motion, $x(\tau)$, constrained so that $x(0) = 0$, $x(t) = 0$ with $x(\tau) > 0$ for $0 < \tau < t$. A Brownian meander on the interval $[0, t]$ is the same except there is no constraint on the value of $x(t)$, other than it is positive. A reflected Brownian bridge on the interval $[0, t]$ is defined as the absolute value $|x(\tau)|$ of a Brownian motion constrained such that $x(0) = 0$ and $x(t) = 0$. For the excursion, on basic dimensional grounds one has that $P(t_m) = t^{-1} f(t_m / t)$, where the function $f(x)$ satisfies the normalisation condition $\int_0^1 f(x)\, dx = 1$. Similarly, in relation to the meander and reflected bridge one can define corresponding functions $g(x)$ and $h(x)$ which are likewise normalised. It follows that the interval length $t$ is only a trivial scaling factor and one can interpret the scaling functions $f(x)$, $g(x)$ and $h(x)$ as being the relevant probability densities for the respective motions on the interval $[0, 1]$. Our primary aim is to compute these functions explicitly. For convenience, we summarise the main findings here;

$$f(x) = 3 \sum_{m,n=1}^{\infty} (-1)^{m+n} \frac{m^2 n^2}{[n^2 x + m^2 (1-x)]^{5/2}} \qquad \text{Excursion} \qquad (1)$$

$$g(x) = 2 \sum_{m=0, n=1}^{\infty} (-1)^{n+1} \frac{n^2}{[n^2 x + (2m+1)^2 (1-x)]^{3/2}} \qquad \text{Meander} \qquad (2)$$

$$h(x) = 2 \sum_{m,n=0}^{\infty} (-1)^{m+n} \frac{(2m+1)(2n+1)}{[(2n+1)^2 x + (2m+1)^2 (1-x)]^{3/2}} \qquad \text{Reflected bridge.} \quad (3)$$

The primary method we employ to derive these results is based on a path-integral technique (in essence, the Feynman-Kac formula). In section 2 we describe how the approach leads naturally to expressions for the joint probability density $P(M, t_m)$. By subsequently integrating over $M$ one can then obtain expressions for the marginal densities $P(t_m)$, and hence obtain the functions $f(x)$, $g(x)$ and $h(x)$ defined above. In section 3 we concentrate on analysing $f(x)$ and $g(x)$ in terms of their moments and asymptotic tails, and show that the results are in excellent accord with numerical simulations. In section 4 we return to the results obtained for $P(M,t_m)$ to show that they may also be obtained by considering certain probabilistic laws known as 'agreement formulae'. These laws are associated with random variables which have been defined and analysed by probability theorists in the study of Brownian motion processes. Finally, in section 5, overall conclusions are drawn.

**2. Deriving the probability densities**

The basic approach outlined in this section has been described in detail in [6,7]. The central idea is that the probability measure $P[x(\tau)]$ associated with an unconstrained Brownian path $x(\tau)$ over the time interval $0 \leq \tau \leq t$ satisfies

$$P[x(\tau)] \propto \exp\left[-\frac{1}{2}\int_0^t \left(\frac{dx}{d\tau}\right)^2 d\tau\right]. \tag{4}$$

From this observation one can systematically construct solutions to the problems of interest by interpreting the 'Lagrangian path-integral' formalism of (4) in terms of an equivalent 'Hamiltonian propagator' formalism within which the constraints on the motion may be accommodated quite naturally. We proceed on a case-by-case basis.

*2.1 Brownian excursion*

Let us first consider the case of the Brownian excursion. With reference to Figure 1, we are interested in those paths $x(\tau)$ which have $x(0)=0$ and $x(t)=0$ with $x(\tau)>0$ for $0<\tau<t$. Note that a continuous-time Brownian motion, starting at $x(0)=0$ at $\tau=0$, will recross the origin an infinite number of times in the time interval $[0,\delta]$ for all $\delta>0$ [10]. Hence it is impossible to maintain the constraint $x(\tau)>0$ for $\tau>0$ if we insist that $x(0)=0$ from the outset. This problem can be circumvented by the following procedure [6,7]. We assume that the process starts at $x(0)=x_0>0$, then impose the constraint $x(\tau)>0$ for $\tau>0$ without any problem, and only take the limit $x_0 \to 0$ at the appropriate stage in the calculation. Similarly, we assume the process ends at $x(t) = x_t > 0$, eventually taking the limit $x_t \to 0$. For computational convenience one can set $x_t = x_0$. Next, one considers the time $t_m$ at which the excursion reaches its (almost surely unique) maximum $M$. Again, we treat this as a limiting process by fixing the value of $x(t_m)$ to be $M - \varepsilon$ whilst imposing the constraint that the actual maximum is less than $M$, with the limit $\varepsilon \to 0$ only being taken at the appropriate stage. With these caveats, and with $t_m$ and $M$ assumed fixed, one can decompose a given path $x(\tau)$ into a left hand segment, for which $0 \leq \tau \leq t_m$, and a right hand segment, for which $t_m \leq \tau \leq t$, wherein for both $0 < x(\tau) < M$ almost surely (see Figure 1). The statistical weight of, say, the left hand segment is proportional to the propagator $\langle x_0 | e^{-\hat{H} t_m} | M - \varepsilon \rangle$, where the Hamiltonian $\hat{H} = -\frac{1}{2}\frac{\partial^2}{\partial x^2} + V(x)$. The potential $V(x)$ has infinite barriers at $x = 0$ and $x = M$; this ensures that the process is constrained to satisfy $0 \leq x(\tau) \leq M$ for $0 \leq \tau \leq t$. The normalised eigenfunctions of $\hat{H}$ are simply $\psi_n(x) = \sqrt{\frac{2}{M}} \sin(\frac{n\pi x}{M})$, whilst the corresponding eigenvalues are given by $E_n = n^2\pi^2/2M^2$. One can easily evaluate the propagator in this eigenbasis;

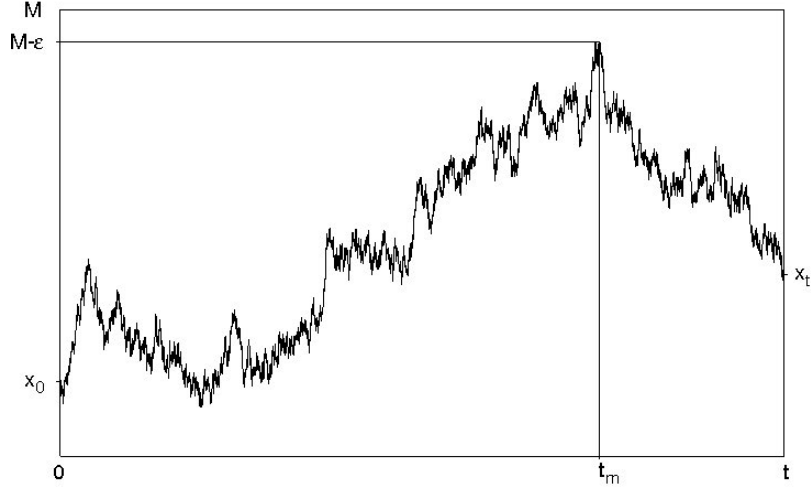

Figure 1: Schematic showing a typical Brownian path $x(\tau)$ constrained so that $x(0) = x_0$, $x(t) = x_t$ and $x(t_m) = M - \varepsilon$, with $0 < x(\tau) < M$ almost surely for $0 \leq \tau \leq t$.

$$\langle x_0 | e^{-\hat{H} t_m} | M - \varepsilon \rangle = \frac{2}{M} \sum_{n=1}^{\infty} \sin\left(\frac{n\pi x_0}{M}\right) \sin\left(\frac{n\pi (M-\varepsilon)}{M}\right) e^{-\frac{n^2\pi^2}{2M^2} t_m}. \quad (5)$$

Similarly, the statistical weight of the right hand segment is proportional to the propagator $\langle M - \varepsilon | e^{-\hat{H}(t-t_m)} | x_0 \rangle$, which may be written down by analogy. With $t_m$ and $M$ fixed, the Markovian nature of the Brownian motion process means that the statistical weight of the right hand segment is independent of the statistical weight of the left hand segment. It follows that the joint probability density, $P(M, t_m)$, after taking the limits $\varepsilon \to 0$ and $x_0 \to 0$, satisfies

$$P(M, t_m) = A \frac{4\pi^4 \varepsilon^2 x_0^2}{M^6} \sum_{m,n=1}^{\infty} (-1)^{m+n} m^2 n^2 e^{-\frac{n^2\pi^2}{2M^2} t_m - \frac{m^2\pi^2}{2M^2}(t-t_m)} + \ldots \quad (6)$$

where … denotes the higher order terms in $\varepsilon$ and $x_0$. The amplitude (i.e. constant of proportionality) $A$, which is a function of $\varepsilon$ and $x_0$, may be determined by the normalisation condition $\int_0^{\infty} \int_0^t P(M, t_m) dt_m dM = 1$. The integrals are straightforward to evaluate and to complete the calculation we make use of the following results;

$$\lim_{\alpha \to -1} \sum_{m,n=1}^{\infty} \alpha^{m+n} \frac{m+n}{mn} = \ln 2; \quad \lim_{\alpha \to -1} \sum_{m,n=1}^{\infty} \alpha^{m+n} \frac{1}{m+n} = \ln 2 - \frac{1}{2}. \quad (7)$$

These results are simple to derive by appropriately differentiating or integrating the basic geometric series $\sum_{n=1}^{\infty} \alpha^n = \alpha/(1-\alpha)$ for $|\alpha| < 1$. This idea of introducing $\alpha$ and letting $\alpha \to -1$ is not just a useful computational aid; it is an important regularisation procedure which gives a precise meaning to certain sums which arise in the analysis. With the help of (7) one therefore finds that, for a Brownian excursion, the joint probability density is given by,

$$P(M,t_m) = \sqrt{2}\,\frac{\pi^{9/2}t^{3/2}}{M^6}\sum_{m,n=1}^{\infty}(-1)^{m+n}m^2 n^2 e^{-\frac{n^2\pi^2}{2M^2}t_m - \frac{m^2\pi^2}{2M^2}(t-t_m)}. \tag{8}$$

We are primarily interested in this paper in the marginal density $P(t_m)$. This may be obtained by integrating (8) over $M$, i.e. $P(t_m) = \int_0^{\infty} P(M,t_m)dM$. Before doing so, however, we make a brief detour by considering the other marginal density associated with the maximum of a Brownian excursion, namely $P(M) = \int_0^t P(M,t_m)dt_m$. This is well known in the literature [15–17];

$$P(M) = \sqrt{2}\pi^{5/2}t^{3/2}\frac{d}{dM}\left\{\frac{1}{M^3}\sum_{n=1}^{\infty}n^2 e^{-\frac{n^2\pi^2}{2M^2}t}\right\} \tag{9}$$

with moments given by,

$$\langle M^k \rangle = \frac{k(k-1)t^{k/2}}{2^{k/2}}\Gamma\left(\frac{k}{2}\right)\zeta(k) \tag{10}$$

where $\zeta(k)$ is the Riemann zeta function. By integrating (8) over $t_m$ one should obtain an expression for $P(M)$ which is equivalent to (9). Interestingly, by doing so one obtains a representation which is quite different;

$$P(M) = \frac{2^{3/2}\pi^{5/2}t^{3/2}}{M^4}\sum_{m,n=1}^{\infty}(-1)^{m+n}\frac{m^2 n^2}{m^2-n^2}\left[e^{-\frac{n^2\pi^2}{2M^2}t} - e^{-\frac{m^2\pi^2}{2M^2}t}\right]. \tag{11}$$

It is by no means obvious that (9) and (11) are equivalent, but in the Appendix we shall prove that this is the case. It follows that the moments calculated using (11) must agree with (10). One quickly establishes that $\langle M \rangle = \sqrt{\pi t/2}$ and $\langle M^2 \rangle = \pi^2 t/6$, as required, and with a little more effort one can also verify that $\langle M^4 \rangle = \pi^4 t^2/30$. For the third moment, however, using (11) and comparing with (10) one obtains (when suitably regularised) an unusual and interesting identity,

$$\lim_{\alpha \to -1} \sum_{m,n=1}^{\infty} \alpha^{m+n} \frac{m^2 n^2}{m+n} \left( \frac{\ln m - \ln n}{m-n} \right) = \frac{3\zeta(3)}{8\pi^2} \tag{12}$$

where $\zeta(3)$ is Apéry's constant. We have checked this numerically to high precision and it is correct, although a *direct* proof of (12) eludes us.

Let us now return to considering the marginal density $P(t_m)$ for a Brownian excursion. By integrating (8) over $M$ one obtains;

$$P(t_m) = 3t^{3/2} \sum_{m,n=1}^{\infty} (-1)^{m+n} \frac{m^2 n^2}{\left[ n^2 t_m + m^2 (t - t_m) \right]^{5/2}} \equiv \frac{1}{t} f\left( \frac{t_m}{t} \right) \tag{13}$$

where the scaling function $f(x)$ is given by,

$$f(x) = 3 \sum_{m,n=1}^{\infty} (-1)^{m+n} \frac{m^2 n^2}{\left[ n^2 x + m^2 (1-x) \right]^{5/2}}. \tag{14}$$

This is our first main result. Specifically, $f(x)$ is the probability density for the time to reach maximum for a Brownian excursion on the interval $[0,1]$. One may easily check that (14) is correctly normalised over this interval, i.e. $\int_0^1 f(x)\, dx = 1$. Thus,

$$\int_0^1 f(x)\, dx = 3 \lim_{\alpha \to -1} \sum_{m,n=1}^{\infty} \alpha^{m+n} m^2 n^2 \int_0^1 \frac{dx}{\left[ n^2 x + m^2 (1-x) \right]^{5/2}}$$

$$= 2 \lim_{\alpha \to -1} \sum_{m,n=1}^{\infty} \alpha^{m+n} \frac{m^2 + mn + n^2}{mn(m+n)} = 1 \tag{15}$$

where we have used the results in (7) when making the final step. In the next section we will consider the low-order moments and asymptotics of $f(x)$, and also make comparison with numerical simulations.

*2.2 Brownian Meander*

Turning now to the case of a Brownian meander, it is straightforward to adapt the above analysis for the excursion and we therefore only present the outline details. The key difference between the meander and the excursion is that there is no constraint on the final co-ordinate of the motion, $x_t$, other than $0 \leq x_t \leq M$ (see Figure 1). Thus one must integrate over this co-ordinate using the result,

$$\int_0^M \sin\left(\frac{m\pi x_t}{M}\right) dx_t = \frac{M}{m\pi}\left[1 - (-1)^m\right]. \tag{16}$$

Proceeding exactly as before one then derives in the limit $\varepsilon \to 0$ and $x_0 \to 0$,

$$P(M, t_m) = B \frac{4\pi^2 \varepsilon^2 x_0}{M^4} \sum_{m,n=1}^{\infty} \left[(-1)^{m+n} - (-1)^n\right] n^2 e^{-\frac{n^2\pi^2}{2M^2}t_m - \frac{m^2\pi^2}{2M^2}(t-t_m)} + \ldots. \tag{17}$$

Again the unknown amplitude $B$ may be determined by the normalisation condition $\int_0^\infty \int_0^t P(M, t_m) dt_m dM = 1$. A useful result in this regard is,

$$\lim_{\alpha \to -1} \sum_{m,n=1}^{\infty} \alpha^n \frac{n}{m(m+n)} = \lim_{\alpha \to -1} \sum_{m,n=1}^{\infty} \alpha^n \left[\frac{1}{m} - \frac{1}{m+n}\right] = -\frac{1}{2}\ln 2. \tag{18}$$

This may be proved by representing the sum in (18) as an integral by first using the identities $m^{-1} \equiv \int_0^\infty e^{-my} dy$ and $(m+n)^{-1} \equiv \int_0^\infty e^{-(m+n)y} dy$ and then interchanging the order of summation and integration. The final result for the joint probability density for a Brownian meander is given by,

$$P(M,t_m) = \frac{\sqrt{2}\pi^{5/2}t^{1/2}}{M^4} \sum_{m,n=1}^{\infty} \left[(-1)^{m+n} - (-1)^n\right] n^2 e^{-\frac{n^2\pi^2}{2M^2}t_m - \frac{m^2\pi^2}{2M^2}(t-t_m)}. \qquad (19)$$

From this result, one can obtain in passing an expression for the marginal density $P(M)$ by integrating over $t_m$;

$$P(M) = \frac{2^{3/2}\pi^{1/2}t^{1/2}}{M^2} \sum_{m,n=1}^{\infty} \left[(-1)^{m+n} - (-1)^n\right] \frac{n^2}{m^2-n^2} \left[ e^{-\frac{n^2\pi^2}{2M^2}t} - e^{-\frac{m^2\pi^2}{2M^2}t} \right]. \qquad (20)$$

There is a standard, well known expression for $P(M)$, namely [18,19],

$$P(M) = 2^{3/2}\sqrt{\pi}t^{1/2} \frac{d}{dM}\left\{ \frac{1}{M} \sum_{n=0}^{\infty} e^{-\frac{(2n+1)^2\pi^2}{2M^2}t} \right\} \qquad (21)$$

with moments [20]

$$\left\langle M^k \right\rangle = k 2^{k/2}(1-2^{1-k}) t^{k/2} \Gamma\!\left(\frac{k}{2}\right) \zeta(k). \qquad (22)$$

As before, we prove in the Appendix that the two representations (20) and (21) are equivalent, as indeed they must be. Using (20) one can determine that the second moment is given by $\left\langle M^2 \right\rangle = \pi^2 t/3$, which agrees with (22). More interestingly, using (20) to determine the first moment and comparing with the result given by (22), namely $\left\langle M \right\rangle = \sqrt{2\pi t}\ln 2$, one obtains the following identity;

$$\lim_{\alpha \to -1} \sum_{m,n=1}^{\infty} \left[\alpha^{m+n} - \alpha^n\right] \frac{n^2}{m+n}\left(\frac{\ln m - \ln n}{m-n}\right) = \frac{1}{2}\ln 2. \qquad (23)$$

Again, we have not been able to prove this result *directly*, but we have checked it numerically to high precision and are satisfied that it is correct.

Returning to the main theme, to obtain the marginal density $P(t_m)$ for a Brownian meander we integrate (19) over $M$ to give;

$$P(t_m) = t^{1/2} \sum_{m,n=1}^{\infty} [(-1)^{m+n} - (-1)^n] \frac{n^2}{[n^2 t_m + m^2(t - t_m)]^{3/2}} \equiv \frac{1}{t} g\left(\frac{t_m}{t}\right) \qquad (24)$$

where

$$g(x) = \sum_{m,n=1}^{\infty} [(-1)^{m+n} - (-1)^n] \frac{n^2}{[n^2 x + m^2(1-x)]^{3/2}}. \qquad (25)$$

One can write this in a slightly neater form by noting that only the terms where $m$ is odd contribute, i.e. $(-1)^{m+n} - (-1)^n = 0$ when $m$ is even. Thus,

$$g(x) = 2 \sum_{m=0, n=1}^{\infty} (-1)^{n+1} \frac{n^2}{[n^2 x + (2m+1)^2(1-x)]^{3/2}}. \qquad (26)$$

This is our second main result. Specifically, $g(x)$ is the probability density for the time to reach maximum for a Brownian meander on the interval $[0,1]$. Again it is useful to check using (25) that $g(x)$ is correctly normalised;

$$\int_0^1 g(x)\,dx = \lim_{\alpha \to -1} \sum_{m,n=1}^{\infty} [\alpha^{m+n} - \alpha^n] n^2 \int_0^1 \frac{dx}{[n^2 x + m^2(1-x)]^{3/2}}$$

$$= 2 \lim_{\alpha \to -1} \sum_{m,n=1}^{\infty} [\alpha^{m+n} - \alpha^n] \frac{n}{m(m+n)} = 1 \qquad (27)$$

where we have used the results in (7) and (18) in the final step. In the next section we will consider the low-order moments and asymptotics of $g(x)$, together with $f(x)$, with comparisons made against numerical simulations. It is perhaps worth stressing in advance that, unlike the function $f(x)$ which is *symmetric* about $x=1/2$, the function $g(x)$ is manifestly *asymmetric*; indeed it diverges as $x \to 1$.

*2.3 Reflected Brownian bridge*

One can easily adapt the above path-integral method to calculate $P(M,t_m)$ for other Brownian motion processes. Anticipating the discussion in section 4, there are good mathematical reasons for studying the reflected Brownian bridge alongside the excursion and meander. The only significant modification to the calculation involves considering a Brownian motion $x(\tau)$ which is constrained to lie in the box $-M < x(\tau) < M$ for $0 \leq \tau \leq t$. We do not give the details of the derivation here, but simply present the results. One finds that,

$$P(M,t_m) = \sqrt{2}\frac{\pi^{5/2} t^{1/2}}{M^4} \sum_{m,n=0}^{\infty} (-1)^{m+n} (m+\tfrac{1}{2})(n+\tfrac{1}{2}) \qquad (28)$$
$$\times e^{-\frac{(n+1/2)^2 \pi^2}{2M^2} t_m - \frac{(m+1/2)^2 \pi^2}{2M^2}(t-t_m)}.$$

By integrating (28) over $t_m$ one therefore has that

$$P(M) = \frac{2^{3/2} \pi^{1/2} t^{1/2}}{M^2} \sum_{m,n=0}^{\infty} (-1)^{m+n} \frac{(m+\tfrac{1}{2})(n+\tfrac{1}{2})}{(m+\tfrac{1}{2})^2 - (n+\tfrac{1}{2})^2} \qquad (29)$$
$$\times \left[ e^{-\frac{(n+1/2)^2 \pi^2}{2M^2} t} - e^{-\frac{(m+1/2)^2 \pi^2}{2M^2} t} \right]$$

which is equivalent (see the Appendix) to the conventional expression [20]

$$P(M) = \sqrt{2\pi t}\, \frac{d}{dM}\left\{ \frac{1}{M} \sum_{n=0}^{\infty} e^{-\frac{(2n+1)^2 \pi^2}{8M^2} t} \right\}. \qquad (30)$$

More pertinently, by integrating (28) over $M$ one obtains the marginal density $P(t_m) = t^{-1} h(t_m / t)$ where the scaling function $h(x)$ is given by,

$$h(x) = 2 \sum_{m,n=0}^{\infty} (-1)^{m+n} \frac{(2m+1)(2n+1)}{\left[(2n+1)^2 x + (2m+1)^2 (1-x)\right]^{3/2}} . \qquad (31)$$

This is our third main result. It is straightforward to check that $h(x)$ is correctly normalised, i.e. $\int_0^1 h(x)\, dx = 1$.

## 3. Analysis and numerical simulations

In this section we concentrate on analysing two of the key results in this paper; the probability density $f(x)$ for the time to reach maximum for a Brownian excursion on the interval $[0,1]$, given by (14), and the probability density $g(x)$ for the time to reach maximum for a Brownian meander on the interval $[0,1]$, given by (26). We begin by considering the excursion. In Figure 2 we plot the function $f(x)$ alongside the results of numerical simulations of the excursion process (based on $10^7$ samples). The results are indistinguishable. We can determine the asymptotic behaviour of $f(x)$ as $x \to 0$ or $x \to 1$ as follows. First we use the identity

$$\frac{1}{a^{5/2}} \equiv \frac{1}{\Gamma(\tfrac{5}{2})} \int_0^{\infty} y^{3/2} e^{-ay} dy \qquad (32)$$

to write $f(x)$ as given by (14) in an equivalent form;

$$f(x) = \frac{4}{\sqrt{\pi}} \int_0^{\infty} y^{3/2} \left\{ \sum_{n=1}^{\infty} (-1)^n n^2 e^{-n^2 x y} \right\} \left\{ \sum_{m=1}^{\infty} (-1)^m m^2 e^{-m^2 (1-x) y} \right\} dy . \qquad (33)$$

By such means one achieves a convenient factorisation, although it comes at a price in that one has to carry out at some stage the parametric integration over the dummy variable $y$. Next,to study the limit $x \to 0$ we substitute $y = z/\sqrt{x}$ in (33) to give

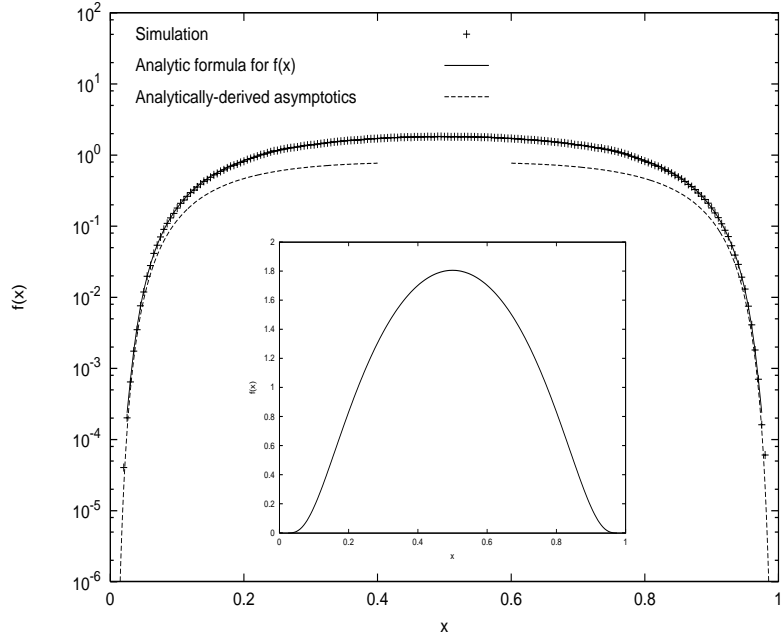

Figure 2: Comparison of simulated results with theoretical predictions for the probability density function $f(x)$. The inset shows the theoretical curve on a linear scale.

$$f(x) = \frac{4}{\sqrt{\pi}} \frac{1}{x^{5/4}} \int_0^\infty z^{3/2} \left\{ \sum_{n=1}^\infty (-1)^n n^2 e^{-n^2 \sqrt{x} z} \right\} \left\{ \sum_{m=1}^\infty (-1)^m m^2 e^{-m^2 (1-x) z/\sqrt{x}} \right\} dz. \quad (34)$$

If we consider the second summation in (34) first, in the limit $x \to 0$ this is dominated by the $m=1$ term, thus,

$$\sum_{m=1}^\infty (-1)^m m^2 e^{-m^2 (1-x) z/\sqrt{x}} \sim -e^{-z/\sqrt{x}}. \quad (35)$$

To deal with the first summation in (34) we use a key identity known from the theory of Jacobi theta functions [21];

$$1 + 2 \sum_{n=1}^\infty (-1)^n e^{-n^2 z} = 2 \left(\frac{\pi}{z}\right)^{1/2} \sum_{n=0}^\infty e^{-\pi^2 (n+1/2)^2 / z}. \quad (36)$$

Differentiating both sides of (36) with respect to $z$, replacing $z \to \sqrt{x}\, z$ and retaining only the dominant term on the right hand side gives the following asymptotic behaviour as $x \to 0$

$$\sum_{n=1}^{\infty} (-1)^n n^2 e^{-n^2 \sqrt{x}\, z} \sim -\frac{\pi^{5/2}}{4 x^{5/4} z^{5/2}} e^{-\pi^2/(4\sqrt{x}\, z)} . \tag{37}$$

This means that by substituting (35) and (37) into (34) one obtains

$$f(x) \sim \frac{\pi^2}{x^{5/2}} \int_0^{\infty} \frac{1}{z} \exp\left[-\frac{1}{\sqrt{x}}\left(\frac{\pi^2}{4z} + z\right)\right] dz . \tag{38}$$

Using Laplace's (saddle point) method to approximate this integral as $x \to 0$ one therefore finds the asymptotic behaviour

$$f(x) \sim 2^{1/2} \frac{\pi^2}{x^{9/4}} e^{-\pi/\sqrt{x}} . \tag{39}$$

This is also plotted in Figure 2, and again the agreement with the numerical simulations is excellent. By symmetry, the asymptotic behaviour of $f(x)$ as $x \to 1$ is given by (39) with the replacement $x \to 1-x$.

One can also consider the moments of $f(x)$. We have calculated the first three moments explicitly using (14) and obtain

$$\langle x \rangle = \frac{1}{2}; \quad \langle x^2 \rangle = \frac{15-\pi^2}{18} = 0.28502...; \quad \langle x^3 \rangle = 1 - \frac{\pi^2}{12} = 0.17753... \tag{40}$$

The first moment follows on the grounds of symmetry, and the third moment may be deduced from the first two moments on the grounds of symmetry. From the simulation results we obtain the numerical estimates: $\langle x \rangle \approx 0.500...$, $\langle x^2 \rangle \approx 0.285...$

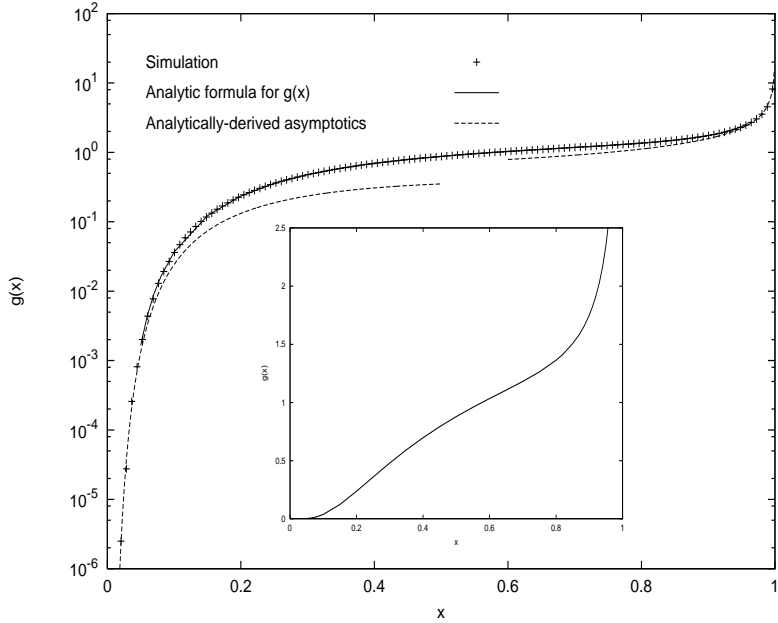

Figure 3: Comparison of simulated results with theoretical predictions for the probability density function $g(x)$. The inset shows the theoretical curve on a linear scale.

and $\langle x^3 \rangle \approx 0.177...$ which are fully consistent. One can calculate higher order moments in principle but this becomes an increasingly laborious task.

Turning now to the case of the meander, in Figure 3 we plot the function $g(x)$ alongside the results of numerical simulations of the meander process (based on $10^7$ samples). Once again the agreement is excellent. This time, the asymptotic behaviour as $x \to 0$ is quite different from the behaviour as $x \to 1$, where $g(x)$ diverges. To investigate these limiting behaviours, we start with (26) and adapt the previous strategy to first rewrite $g(x)$ in the form,

$$g(x) = \frac{4}{\sqrt{\pi}} \int_0^\infty y^{1/2} \left\{ \sum_{n=1}^\infty (-1)^{n+1} n^2 e^{-n^2 x y} \right\} \left\{ \sum_{m=0}^\infty e^{-(2m+1)^2 (1-x) y} \right\} dy . \qquad (41)$$

To study the limit $x \to 0$ we substitute $y = z/\sqrt{x}$ in (41) to give

$$g(x) = \frac{4}{\sqrt{\pi} x^{3/4}} \int_0^\infty z^{1/2} \left\{ \sum_{n=1}^\infty (-1)^{n+1} n^2 e^{-n^2 \sqrt{x} z} \right\} \left\{ \sum_{m=0}^\infty e^{-4(m+1/2)^2 (1-x) z / \sqrt{x}} \right\} dz. \qquad (42)$$

As $x \to 0$, to approximate the first summation in (42) we can use the result (37), whilst for the second summation in (42) we need retain only the $m=0$ term so that

$$\sum_{m=0}^\infty e^{-4(m+1/2)^2 (1-x) z / \sqrt{x}} \sim e^{-z/\sqrt{x}}. \qquad (43)$$

This means that

$$g(x) \sim \frac{\pi^2}{x^2} \int_0^\infty \frac{1}{z^2} \exp\left[ -\frac{1}{\sqrt{x}} \left( \frac{\pi^2}{4z} + z \right) \right] dz. \qquad (44)$$

Once again by using Laplace's method to approximate the integral as $x \to 0$ we obtain the asymptotic behaviour of $g(x)$ for $x \to 0$;

$$g(x) \sim 2^{3/2} \frac{\pi}{x^{7/4}} e^{-\pi/\sqrt{x}}. \qquad (45)$$

This is plotted in Figure 3 and agrees very well with the simulations. Turning now to the limit $x \to 1$, we cannot rely on symmetry arguments as was the case for the excursion, so we substitute $y = z/\sqrt{1-x}$ into (41) to give,

$$g(x) = \frac{4}{\sqrt{\pi} (1-x)^{3/4}} \int_0^\infty z^{1/2} \left\{ \sum_{n=1}^\infty (-1)^{n+1} n^2 e^{-n^2 x z / \sqrt{1-x}} \right\} \left\{ \sum_{m=0}^\infty e^{-4(m+1/2)^2 \sqrt{1-x} z} \right\} dz. \qquad (46)$$

Considering the second summation in (46) first, we can manipulate the fundamental theta function identity (36) to show that in the limit $x \to 1$

$$\sum_{m=0}^{\infty} e^{-4(m+1/2)^2 \sqrt{1-x}\, z} \sim \frac{\sqrt{\pi}}{4}\left(\frac{1}{1-x}\right)^{1/4} \frac{1}{z^{1/2}}. \tag{47}$$

For the first summation in (46) one has to be careful; it turns out that *all* the terms in it contribute to the integral to the same order as $x \to 1$. Thus one has that,

$$g(x) \sim \frac{1}{(1-x)} \int_0^{\infty} \sum_{n=1}^{\infty} (-1)^{n+1} n^2 e^{-n^2 z/\sqrt{1-x}}\, dz = \frac{1}{\sqrt{1-x}} \sum_{n=1}^{\infty} (-1)^{n+1}. \tag{48}$$

The summation requires proper regularisation for its correct interpretation, and gives

$$\sum_{n=1}^{\infty} (-1)^{n+1} = \lim_{\alpha \to -1} \sum_{n=1}^{\infty} \alpha^{n+1} = \lim_{\alpha \to -1} \frac{\alpha^2}{1-\alpha} = \frac{1}{2}. \tag{49}$$

Thus the asymptotic behaviour of $g(x)$ as $x \to 1$ turns out to be very simple;

$$g(x) \sim \frac{1}{2\sqrt{1-x}}. \tag{50}$$

Again this accords well with the simulations, as shown in Figure 3. The square root divergence is the same, except for the pre-factor, as that found for the case of unconstrained Brownian motion, as discussed in the introduction.

For completeness, we have also calculated the first two moments of $g(x)$ explicitly using (25), with the result that,

$$\langle x \rangle = \frac{\pi^2 - 4}{8} = 0.73370...; \quad \langle x^2 \rangle = \frac{19\pi^2 - 60}{216} = 0.59038.... \tag{51}$$

From the simulation results we obtain the numerical estimates $\langle x \rangle \approx 0.733...$ and $\langle x^2 \rangle \approx 0.590...$, which again are fully consistent.

For the reflected Brownian bridge, one can follow similar procedures to derive the asymptotic behaviour of the function $h(x)$ and calculate the moments. We omit the details here. The derivation of $h(x)$ presented in the previous section serves primarily to make a connection with what follows.

**4. An alternative derivation via 'agreement formulae'**

The results derived above for the joint probability density $P(M, t_m)$ for the excursion (8), meander (19), and reflected bridge (28) have a particular mathematical structure which can be understood from a different perspective. In this section we show, in outline form only, that these results are manifestations of probabilistic laws associated with three random variables which have previously been studied in the Brownian motion literature; see [11] and, in particular, [22-24] and references therein. Such work builds upon the initial path decomposition work of Williams [25], which in turn provides another interpretation of the findings presented in [12] for the maximum of a Brownian motion up to its first-passage time, and upon the results presented in [26]. By such means one can provide an alternative, although less physically intuitive, method of derivation of the main results in this paper. First consider three variables $S$, $T$ and $C$, which are characterised by the Laplace transform of their respective probability densities $P_S(u)$, $P_T(u)$ and $P_C(u)$ [24, 27];

$$\mathrm{E}\left[e^{-\lambda S}\right] = \frac{\sqrt{2\lambda}}{\sinh(\sqrt{2\lambda})} = 2\sum_{n=1}^{\infty} (-1)^{n+1} \frac{n^2\pi^2}{n^2\pi^2 + 2\lambda} \tag{52}$$

$$\mathrm{E}\left[e^{-\lambda T}\right] = \frac{\tanh(\sqrt{2\lambda})}{\sqrt{2\lambda}} = 2\sum_{n=0}^{\infty} \frac{1}{(n+\tfrac{1}{2})^2\pi^2 + 2\lambda} \tag{53}$$

$$\mathrm{E}\left[e^{-\lambda C}\right] = \frac{1}{\cosh(\sqrt{2\lambda})} = 2\sum_{n=0}^{\infty} (-1)^n \frac{(n+\tfrac{1}{2})\pi}{(n+\tfrac{1}{2})^2\pi^2 + 2\lambda} \tag{54}$$

where $\mathrm{E}[e^{-\lambda S}] \equiv \int_0^\infty P_S(u) e^{-\lambda u} du$ etc. and the series expansions on the right hand side of (52), (53) and (54) are well known identities [28]. From these identities it follows that the probability densities can be written as,

$$P_S(u) = \pi^2 \sum_{n=1}^\infty (-1)^{n+1} n^2 e^{-\frac{n^2 \pi^2}{2} u} \tag{55}$$

$$P_T(u) = \sum_{n=0}^\infty e^{-\frac{(n+1/2)^2 \pi^2}{2} u} \tag{56}$$

$$P_C(u) = \pi \sum_{n=0}^\infty (-1)^n (n+\tfrac{1}{2}) e^{-\frac{(n+1/2)^2 \pi^2}{2} u} . \tag{57}$$

With this preamble, the structure of (8), (19) and (28) can be deduced by considering certain so-called 'agreement formulae' [22-24]. These are identities in law between two-dimensional random variables which relate to various fundamental processes defined on the interval [0,1] and which are valid for an *arbitrary* function $f$ ;

$$\mathrm{E}[f(M^2, t_m)] = \sqrt{\frac{\pi}{2}} \mathrm{E}\left[ f\left(\frac{1}{S+S'}, \frac{S}{S+S'}\right) \sqrt{S+S'} \right] \quad \text{Excursion} \tag{58}$$

$$\mathrm{E}[f(M^2, t_m)] = \sqrt{\frac{\pi}{8}} \mathrm{E}\left[ f\left(\frac{1}{S+\tfrac{1}{4}T}, \frac{S}{S+\tfrac{1}{4}T}\right) \frac{1}{\sqrt{S+\tfrac{1}{4}T}} \right] \quad \text{Meander} \tag{59}$$

$$\mathrm{E}[f(M^2, t_m)] = \sqrt{\frac{\pi}{2}} \mathrm{E}\left[ f\left(\frac{1}{C+C'}, \frac{C}{C+C'}\right) \frac{1}{\sqrt{C+C'}} \right] \quad \text{Reflected bridge.} \tag{60}$$

In these formulae, the variables $S$, $T$ and $C$ are the random variables described above with probability densities given by (55), (56) and (57) respectively, whilst $M$ and $t_m$ have the same meaning (in relation to the named process) as they have had throughout the paper. All three results, namely (58), (59) and (60), are of the same generic form;

$$\mathrm{E}[f(M^2, t_m)] = \alpha \mathrm{E}\left[ f\left(\frac{1}{U+V}, \frac{U}{U+V}\right)(U+V)^{\mu} \right] \qquad (61)$$

where $\alpha$ and $\mu$ are chosen accordingly and, crucially, on the right hand side the random variables $U$ and $V$ are *independent*. The first (58) and third (60) of these results are particular cases of a Bessel bridge process of dimension $d = 2(1+\mu)$ represented in terms of two independent Bessel processes considered up to their first hitting times of 1, and placed 'back-to-back' [23,24]; see [11] for further references. The second result (59) is not precisely found in the literature, but may be obtained as a consequence of (60) (the details will be presented in another publication). For now, if we denote the probability density of $U$ by $h(u)$ and the probability density of $V$ by $k(v)$, one can use (61) to obtain a relationship between $P(M, t_m)$, the joint probability density of $(M, t_m)$, and the pair $(h, k)$. Again we skip the details, but in summary one can show that (61) implies that the function $P(*,*)$ must satisfy

$$\frac{1}{2\alpha} \frac{1}{(u+v)^{\mu+5/2}} P\left(\frac{1}{\sqrt{u+v}}, \frac{u}{u+v}\right) = h(u)k(v). \qquad (62)$$

Letting $u = t_m / M^2$ and $v = (1-t_m)/M^2$, and then exploiting the scaling properties of the processes to consider the general interval $[0,t]$ rather than just the interval $[0,1]$, it follows that the joint probability densities $P(M, t_m)$ for the excursion, meander and reflected bridge are all of the form,

$$P(M, t_m) = 2\alpha \frac{t^{\mu+1}}{M^{2\mu+5}} h\left(\frac{t_m}{M^2}\right) k\left(\frac{t-t_m}{M^2}\right). \qquad (63)$$

With reference to (58), (59) and (60), it therefore follows that

$$P(M,t_m) = \sqrt{2\pi}\,\frac{t^{3/2}}{M^6}\,P_S\!\left(\frac{t_m}{M^2}\right) \times P_S\!\left(\frac{t-t_m}{M^2}\right) \qquad \text{Excursion} \qquad (64)$$

$$P(M,t_m) = \sqrt{\frac{\pi}{2}}\,\frac{t^{1/2}}{M^4}\,P_S\!\left(\frac{t_m}{M^2}\right) \times 4P_T\!\left(\frac{4(t-t_m)}{M^2}\right) \qquad \text{Meander} \qquad (65)$$

$$P(M,t_m) = \sqrt{2\pi}\,\frac{t^{1/2}}{M^4}\,P_C\!\left(\frac{t_m}{M^2}\right) \times P_C\!\left(\frac{t-t_m}{M^2}\right) \qquad \text{Reflected bridge.} \qquad (66)$$

One may easily check using (55), (56) and (57) that (64), (65) and (66) reproduce in full the earlier results derived using the path-integral method. An expanded version of the discussion in this section will be given in a subsequent publication.

## 5. Conclusions

By way of conclusion, we have succeeded in deriving expressions for the probability density $P(t_m)$ for the time to reach maximum for a Brownian excursion (14), a Brownian meander (26) and a reflected Brownian bridge (31). This has been achieved first by using a path-integral technique, suitably adapted to each case in turn, with the key feature of introducing appropriate cut-offs which are then allowed to tend to zero. The derivation is reasonably transparent and, of course, can be adapted to give comparatively simple derivations of the 'arcsine law' and the 'uniform law' mentioned in the introduction. Indeed, this was one of the earliest applications of the Feynman-Kac formula. In passing, we have also derived in each case an expression for the probability density $P(M)$ associated with the distribution of the maximum. The representations for $P(M)$ thus obtained are quite different from the standard representations found in the literature but we have been able to prove their equivalence (see the Appendix). By considering the moments, therefore, this leads to new, non-trivial identities (such as (12) and (23)) which we have verified numerically to high precision. For the excursion and meander, we have also analysed the moments and asymptotic limits of $P(t_m)$ in some detail and demonstrated that the theoretical results are in complete accord with numerical simulations. Finally, the 'agreement formulae' (58), (59) and (60) provide an alternative route to the derivation of the main results. At a fundamental level, this points to fascinating and deep connections with other problems and is a promising avenue for further study.

**Appendix: Establishing the equivalence of certain probability densities**

The expression for the probability density $P(M)$ for the maximum of a Brownian excursion obtained using the path integral method, (9), is quite different from the conventional expression, (11). Here we establish the equivalence. After simplifying both (9) and (11) this is tantamount to having to prove that

$$2\sum_{m,n=1}^{\infty}(-1)^{m+n}\frac{m^2n^2}{m^2-n^2}\left[e^{-n^2x}-e^{-m^2x}\right]\stackrel{?}{=}-3\sum_{n=1}^{\infty}n^2e^{-n^2x}+2x\sum_{n=1}^{\infty}n^4e^{-n^2x}. \quad (A1)$$

It is expedient (temporarily) to separate out the $m=n$ term on the left hand side (which is evaluated by a limiting procedure) since it exactly cancels the second term on the right hand side. By further integrating both sides with respect to $x$ this means that (A1) reduces to

$$2\sum_{\substack{m,n=1\\m\neq n}}^{\infty}\frac{(-1)^{m+n}}{m^2-n^2}\left[n^2e^{-m^2x}-m^2e^{-n^2x}\right]\stackrel{?}{=}3\sum_{n=1}^{\infty}e^{-n^2x}. \quad (A2)$$

Next we take the Laplace transform of both sides of (A2) with respect to $x$ to give,

$$-2\sum_{\substack{m,n=1\\m\neq n}}^{\infty}(-1)^{m+n}\left[\frac{(m^2+n^2+s)}{(m^2+s)(n^2+s)}\right]\stackrel{?}{=}3\sum_{n=1}^{\infty}\frac{1}{n^2+s}. \quad (A3)$$

It is now helpful to 'add back' the $m=n$ term on the left hand side and to the right hand side also. After some simple algebra (A3) then reduces to

$$2\sum_{m,n=1}^{\infty}(-1)^{m+n}\left[\frac{(m^2+n^2+s)}{(m^2+s)(n^2+s)}\right]\stackrel{?}{=}\sum_{n=1}^{\infty}\frac{1}{n^2+s}-2s\sum_{n=1}^{\infty}\frac{1}{(n^2+s)^2}. \quad (A4)$$

One can further simplify the left hand side by writing

$$\frac{(m^2+n^2+s)}{(m^2+s)(n^2+s)} \equiv \frac{1}{(n^2+s)} + \frac{1}{(m^2+s)} - \frac{s}{(m^2+s)(n^2+s)} \tag{A5}$$

whereupon the overall task condenses down to demonstrating that

$$-2\sum_{n=1}^{\infty}\frac{(-1)^n}{n^2+s} - 2s\left(\sum_{n=1}^{\infty}\frac{(-1)^n}{n^2+s}\right)^2 \stackrel{?}{=} \sum_{n=1}^{\infty}\frac{1}{n^2+s} - 2s\sum_{n=1}^{\infty}\frac{1}{(n^2+s)^2}. \tag{A6}$$

The remaining steps needed to establish that (A6) is a rigorous equality simply require one to use the identities [28]

$$\sum_{n=1}^{\infty}\frac{1}{n^2+s} = \frac{\pi}{2\sqrt{s}}\coth(\pi\sqrt{s}) - \frac{1}{2s} \tag{A7}$$

$$\sum_{n=1}^{\infty}\frac{(-1)^n}{n^2+s} = \frac{\pi}{2\sqrt{s}}\operatorname{cosech}(\pi\sqrt{s}) - \frac{1}{2s} \tag{A8}$$

$$\sum_{n=1}^{\infty}\frac{1}{(n^2+s)^2} = -\frac{1}{2s^2} + \frac{\pi}{4s^{3/2}}\coth(\pi\sqrt{s}) + \frac{\pi^2}{4s}\operatorname{cosech}^2(\pi\sqrt{s}) \tag{A9}$$

where (A9) can be deduced from (A7) by differentiating both sides with respect to $s$. It is now elementary to show that the left hand side and the right hand side of (A6) are equal. It follows that since the Laplace transforms of (9) and (11) are equal, then the functions themselves are equal, and the proof is complete.

Similarly for the meander, the expression for the probability density of the maximum obtained using the path integral method, (20), is quite different from the conventional expression, (21). To establish the equivalence we proceed in the same manner as above. First, it is a simple matter to reduce the task to one of proving that

$$\sum_{\substack{m,n=1 \\ m \neq n}}^{\infty} \left[(-1)^{m+n} - (-1)^n\right] \frac{n^2}{m^2 - n^2} \left[e^{-n^2 x} - e^{-m^2 x}\right] \stackrel{?}{=} -\sum_{n=0}^{\infty} e^{-(2n+1)^2 x}. \quad \text{(A10)}$$

Again we take Laplace transforms with respect to $x$ so as to reduce (A10) to

$$\sum_{\substack{m,n=1 \\ m \neq n}}^{\infty} \left[(-1)^{m+n} - (-1)^n\right] \frac{n^2}{(n^2+s)(m^2+s)} \stackrel{?}{=} -\sum_{n=0}^{\infty} \frac{1}{(2n+1)^2+s}. \quad \text{(A11)}$$

By 'adding back' the $m = n$ term one can then reduce (A11) to

$$\sum_{m,n=1}^{\infty} \left[(-1)^{m+n} - (-1)^n\right] \frac{n^2}{(n^2+s)(m^2+s)}$$
$$\stackrel{?}{=} \sum_{n=0}^{\infty} \frac{1}{(2n+1)^2+s} - 2s \sum_{n=0}^{\infty} \frac{1}{((2n+1)^2+s)^2}. \quad \text{(A12)}$$

Further manipulations of the left hand side of (A12) yield additional simplifications until finally the task is to establish the following;

$$\left\{\sum_{n=1}^{\infty} \frac{(-1)^n}{n^2+s}\right\} \left\{\sum_{n=0}^{\infty} \frac{1}{(2n+1)^2+s}\right\} \stackrel{?}{=} -\sum_{n=0}^{\infty} \frac{1}{((2n+1)^2+s)^2}. \quad \text{(A13)}$$

This is easily done using the results (A7) and (A8). Thus by subtracting (A8) from (A7) one obtains

$$\sum_{n=0}^{\infty} \frac{1}{(2n+1)^2+s} = \frac{\pi}{4\sqrt{s}} \left[\coth(\pi\sqrt{s}) - \text{cosech}(\pi\sqrt{s})\right] = \frac{\pi}{4\sqrt{s}} \tanh\left(\frac{\pi\sqrt{s}}{2}\right). \quad \text{(A14)}$$

By further differentiating (A14) with respect to $s$ one obtains

$$\sum_{n=0}^{\infty}\frac{1}{((2n+1)^2+s)^2} = \frac{\pi}{8s^{3/2}}\left[\coth(\pi\sqrt{s})-\operatorname{cosech}(\pi\sqrt{s})\right] \\ +\frac{\pi^2}{8s}\operatorname{cosech}^2(\pi\sqrt{s})\left[1-\cosh(\pi\sqrt{s})\right]$$ (A15)

The proof then follows by direct substitution.

Finally, for the reflected bridge, to establish the equivalence between (29) and (30) one can follow the above procedure, which requires one to show that

$$2\sum_{\substack{m,n=0\\m\neq n}}^{\infty}(-1)^{m+n}\frac{(2m+1)(2n+1)}{((2m+1)^2+s)((2n+1)^2+s)}\stackrel{?}{=}-\sum_{n=0}^{\infty}\frac{1}{(2n+1)^2+s}$$ (A16)

which is the same as showing that

$$2\left[\sum_{n=0}^{\infty}(-1)^n\frac{(2n+1)}{(2n+1)^2+s}\right]^2 \stackrel{?}{=} -\sum_{n=0}^{\infty}\frac{1}{(2n+1)^2+s}+2\sum_{n=0}^{\infty}\frac{(2n+1)^2}{((2n+1)^2+s)^2}.$$ (A17)

The right hand side of this expression may be rewritten using (A14) and (A15). The left hand side may then be shown to be equivalent using the result [28]

$$\sum_{n=0}^{\infty}(-1)^n\frac{(2n+1)}{(2n+1)^2+s} = \frac{\pi}{4}\operatorname{sech}\left(\frac{\pi\sqrt{s}}{2}\right).$$ (A18)